\documentclass[twocolumn,prl]{revtex4}
\usepackage[T1]{fontenc}
\usepackage{amsmath,amsbsy,amssymb,graphicx}
\usepackage{times}

\let\mathbf=\boldsymbol

\begin{document}

\title{{\Large Valley-Polarized Metals and Quantum Anomalous Hall Effect in
Silicene}}
\author{Motohiko Ezawa}
\affiliation{Department of Applied Physics, University of Tokyo, Hongo 7-3-1, 113-8656,
Japan }

\begin{abstract}
Silicene is a monolayer of silicon atoms forming a two-dimensional honeycomb
lattice, which shares almost every remarkable property with graphene. The
low energy structure of silicene is described by Dirac electrons with
relatively large spin-orbit interactions due to its buckled structure. The
key observation is that the band structure is controllable by applying
electric field to silicene. We explore the phase diagram of silicene
together with exchange field $M$ and by applying electric field $E_{z}$.
There appear quantum anomalous Hall (QAH) insulator, valley polarized metal
(VPM), marginal valley polarized metal (M-VPM), quantum spin Hall (QSH)
insulator and band insulator (BI). They are characterized by the Chern
numbers and/or by the edge modes of a nanoribbon. It is intriguing that
electrons have been moved from a conduction band at the K point to a valence
band at the K' point for $E_{z}>0$ in the VPM. We find in the QAH phase that
almost flat gapless edge modes emerge and that spins form a momentum-space
skyrmion to yield the Chern number. It is remarkable that a topological
quantum phase transition can be induced simply by changing electric field in
a single silicene sheet.
\end{abstract}

\maketitle


Silicene, a monolayer of silicon atoms forming a two-dimensional honeycomb
lattice, has been synthesized\cite{Lalmi,GLayPRL,Takagi} and attracts much
attention\cite{Shiraishi,LiuPRL,LiuPRB,EzawaNJP,EzawaJ} recently. Almost
every striking property of graphene could be transferred to this innovative
material. It has additionally a salient feature, that is a buckled structure%
\cite{Shiraishi,LiuPRL} owing to a large ionic radius of silicon. Silicene
has a relatively large spin-orbit (SO) gap of $1.55$meV, which provides a
mass to Dirac electrons. Furthermore, we may control experimentally the mass%
\cite{EzawaNJP} by applying the electric field $E_{z}$. Silicene undergoes a
topological phase transition from a quantum spin Hall (QSH) state to a band
insulator (BI) as $|E_{z}|$ increases\cite{EzawaNJP}. A QSH state is
characterized by a full insulating gap in the bulk and \textit{helical}
gapless edges\cite{KaneMele,Wu,Hasan,Qi}.

There exits another state of matter in graphene\cite{Qiao,Tse,Yang}, that is
a quantum anomalous Hall (QAH) state\cite{Liu,Onoda}, characterized by a
full insulating gap in the bulk and \textit{chiral} gapless edges. Unlike
the quantum Hall effect, which arises from Landau-level quantization in a
strong magnetic field, the QAH effect is induced by internal magnetization
and SO coupling.

In this paper we analyze the band structure of silicene together with
exchange field $M$ and by applying electric field $E_{z}$ to silicene. We
explore the phase diagram in the $E_{z}$-$M$ plane. Silicene has a rich
varieties of phases because the electric field $E_{z}$ and the exchange
field $M$ have different effects on the conduction and valence bands
characterized by the spin and valley indices. There are insulator phases,
which are the QSH, QAH and BI phases. There emerges a new type of metal
phase, the valley-polarized metal (VPM) phase, where electrons have been
moved from a conduction band at the K point to a valence band at the K'
point for $E_{z}>0$. Such a phase is utterly unknown in literature as far as
we are aware of. There are also metallic states on phase boundaries, which
are metal (M), marginal-VPM (M-VPM) and spin VPM (SVPM) states. All these
phases and states are characterized by the Chern numbers and/or by the edge
modes of a nanoribbon. It is possible to materialize any one of them by
controlling $E_{z}$ at an appropriate value of $M$. Furthermore, as we have
pointed out elsewhere\cite{EzawaNJP}, by applying an inhomogeneous field $%
E_{z}$, it is possible to materialize some of these topological phases
together with states on the phase boundaries simultaneously in a single
silicene sheet.

Silicene consists of a honeycomb lattice of silicon atoms with two
sublattices made of A sites and B sites. The states near the Fermi energy
are $\pi $ orbitals residing near the K and K' points at opposite corners of
the hexagonal Brillouin zone. We refer to the K or K' point also as the K$%
_{\eta }$ point with the valley index $\eta =\pm 1$. We take a silicene
sheet on the $xy$-plane, and apply the electric field $E_{z}$ perpendicular
to the plane. Due to the buckled structure the two sublattice planes are
separated by a distance, which we denote by $2\ell $ with $\ell =0.23$\AA .
It generates a staggered sublattice potential $\varpropto 2\ell E_{z}$
between silicon atoms at A sites and B sites.

The silicene system is described by the four-band second-nearest-neighbor
tight binding model,

\begin{align}
H& =-t\sum_{\left\langle i,j\right\rangle \alpha }c_{i\alpha }^{\dagger
}c_{j\alpha }+i\frac{\lambda _{\text{SO}}}{3\sqrt{3}}\sum_{\left\langle
\!\left\langle i,j\right\rangle \!\right\rangle \alpha \beta }\nu
_{ij}c_{i\alpha }^{\dagger }\sigma _{\alpha \beta }^{z}c_{j\beta }  \notag \\
& +i\lambda _{\text{R1}}(E_{z})\sum_{\left\langle i,j\right\rangle \alpha
\beta }c_{i\alpha }^{\dagger }\left( \mathbf{\sigma }\times \hat{\mathbf{d}}%
_{ij}\right) _{\alpha \beta }^{z}c_{j\beta }  \notag \\
& -i\frac{2}{3}\lambda _{\text{R2}}\sum_{\left\langle \!\left\langle
i,j\right\rangle \!\right\rangle \alpha \beta }\mu _{i}c_{i\alpha }^{\dagger
}\left( \mathbf{\sigma }\times \hat{\mathbf{d}}_{ij}\right) _{\alpha \beta
}^{z}c_{j\beta }  \notag \\
& +\ell \sum_{i\alpha }\mu _{i}E_{z}c_{i\alpha }^{\dagger }c_{i\alpha
}+M\sum_{i\alpha }c_{i\alpha }^{\dagger }\sigma _{z}c_{i\alpha },
\label{BasicHamil}
\end{align}%
where $c_{i\alpha }^{\dagger }$ creates an electron with spin polarization $%
\alpha $ at site $i$, and $\left\langle i,j\right\rangle /\left\langle
\!\left\langle i,j\right\rangle \!\right\rangle $ run over all the
nearest/next-nearest neighbor hopping sites. We explain each term. (i) The
first term represents the usual nearest-neighbor hopping with the transfer
energy $t=1.6$eV. (ii) The second term represents the effective SO coupling
with $\lambda _{\text{SO}}=3.9$meV, where $\mathbf{\sigma }=(\sigma
_{x},\sigma _{y},\sigma _{z})$ is the Pauli matrix of spin, with $\nu
_{ij}=+1$ if the next-nearest-neighboring hopping is anticlockwise and $\nu
_{ij}=-1$ if it is clockwise with respect to the positive $z$ axis. (iii)
The third term represents the first Rashba SO coupling associated with the
nearest neighbor hopping, which is induced by external electric field\cite%
{Hongki,Tse}. It satisfies $\lambda _{\text{R1}}(0)=0$ and becomes of the order of $10\mu $eV
at the critical electric field $E_{\text{c}}=\lambda _{\text{SO}}/\ell =17$%
meV\AA $^{-1}$. (iv) The forth term represents the second Rashba SO coupling
with $\lambda _{\text{R2}}=0.7$meV associated with the next-nearest neighbor
hopping term, where $\mu _{i}=\pm 1$ for the A (B) site, and $\hat{\mathbf{d}%
}_{ij}=\mathbf{d}_{ij}/\left\vert \mathbf{d}_{ij}\right\vert $ with the
vector $\mathbf{d}_{ij}$ connecting two sites $i$ and $j$ in the same
sublattice. (v) The fifth term is the staggered sublattice potential term.
(vi) The sixth term represents the exchange magnetization: Exchange field $M$
may arise due to proximity coupling to a ferromagnet such as depositing Fe
atoms to the silicene surface or depositing silicene to a ferromagnetic
insulating substrate, as has been argued for graphene\cite{Qiao,Tse,Yang}.
The Hamiltonian (\ref{BasicHamil}) can also be used to describe germanene,
which is a honeycomb structure of germanium\cite{LiuPRL,LiuPRB}, where
various parameters are $t=1.3$eV, $\lambda _{\text{SO}}=43$meV, $\lambda _{%
\text{R2}}=10.7$meV and $\ell =0.33$\AA .

\begin{figure}[t]
\centerline{\includegraphics[width=0.34\textwidth]{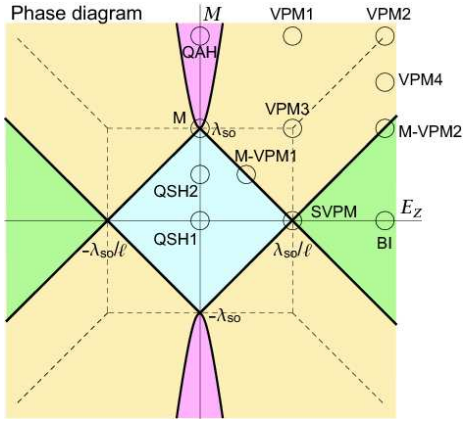}}
\caption{(Color online) Phase diagram in the $E_{z}$-$M$ plane. Heavy lines
represent phase boundaries, where the system becomes metallic. Chern and
spin-Chern numbers ($C,C_{s}$) are well defined and given in insulator
phases. Dotted lines represent the points where the band gap closes, which
are within the VPM phase. A circle shows a point where the energy spectrum
is calculated and shown in Fig.\protect\ref{FigRBand}.}
\label{FigPhaseDB}
\end{figure}

In this paper we derive the topological phase diagram in the $E_{z}$-$M$
plane and make its physical interpretation. The topological quantum numbers
are the Chern number $C$ and the $\mathbb{Z}_{2}$ index. If the spin $s_{z}$
is a good quantum number, the $\mathbb{Z}_{2}$ index is identical to the
spin-Chern number $C_{s}$. They are defined when the state is gapped and
when the Fermi level is taken within the gap, and given by $C=C_{+}+C_{-}$
and $C_{s}=\frac{1}{2}(C_{+}-C_{-})$, where $C_{\pm }$ is the summation of
the Berry curvature in momentum space over all occupied states of electrons
with $s_{z}=\pm 1$. They are well defined even if the spin is not a good
quantum number\cite{Prodan09B,Yang}. In the present model the spin is not a
good quantum number because of spin mixing due to the Rashba couplings $%
\lambda _{\text{R1}}$ and $\lambda _{\text{R2}}$, and the resulting angular
momentum eigenstates are indexed by the spin chirality $s=\pm 1$. We can
calculate these numbers at each point in the $E_{z}$-$M$ plane by using the
standard formulas\cite{Qiao,Tse,Yang}.

\begin{figure}[t]
\centerline{\includegraphics[width=0.50\textwidth]{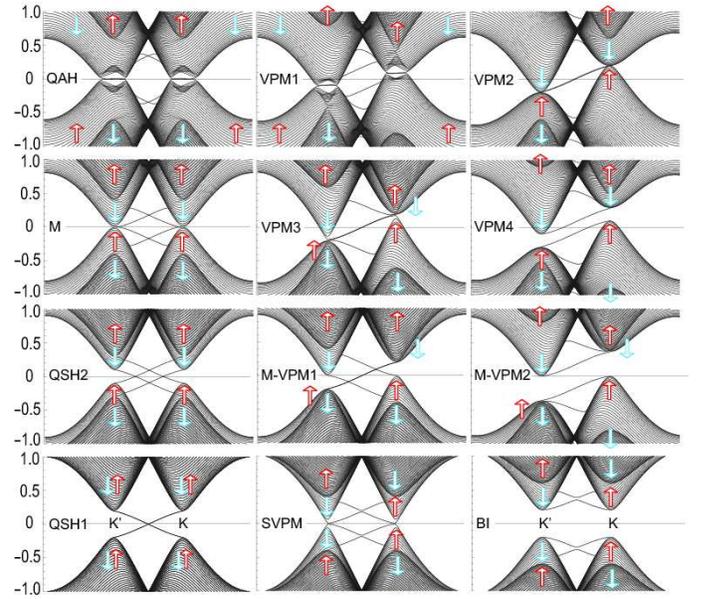}}
\caption{(Color online) The band structure of a silicene nanoribbon at
marked points in the phase diagram (Fig.\protect\ref{FigPhaseDB}). The
vertical axis is the energy in uit of $t$, and the horizontal axis is the
momentum. We can clearly see the Dirac cones representing the energy
spectrum of the bulk, as explained in Fig.\protect\ref{FigQAH}. Lines
connecting the two Dirac cones are edge modes. The spin $s_{z}$ is
practically a good quantum number, which we have assigned to the Dirac
cones. An enlaraged figure of the QAH state is given in Fig.\protect\ref%
{FigQAH}.}
\label{FigRBand}
\end{figure}

We present our result on the phase diagram in Fig.\ref{FigPhaseDB}. We show
later how to derive the phase boundaries based on the low-energy Dirac
theory. We have also calculated the band structure of a silicene nanoribbon
with zigzag edges, which we give in Fig.\ref{FigRBand} for typical points in
the phase diagram. The topological numbers are $(C,C_{s})=(0,0)$ in the BI
phase, $(0,1)$ in the QSH phase, $(2,0)$ in the QAH phase with $M>0$ and $%
(-2,0)$ in the QAH phase with $M<0$. In all these states the band gap is
open, where the Fermi level is present, and they are insulators.

\begin{figure}[t]
\centerline{\includegraphics[width=0.48\textwidth]{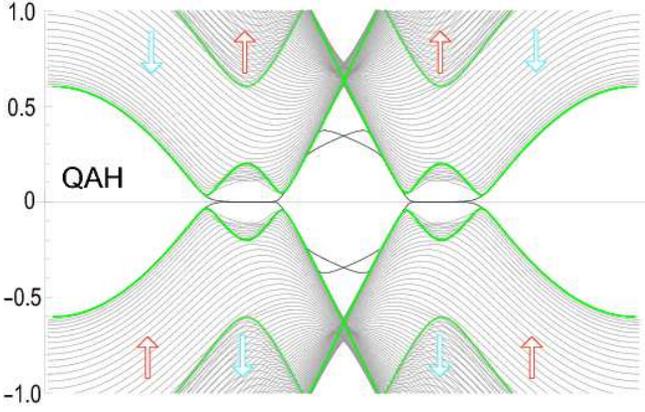}}
\caption{(Color online) The energy spectrum of a QAH state. Gray curves are
for a nanoribbon, which are identical to Fig.\protect\ref{FigRBand}(QAH).
Heavy green curves represent the energy spectrum of the bulk, calculated
independently. A gap opens in the bulk spectrum, where almost flat gapless
modes appear at the edges of a nanoribbon. A red (blue) arrow indicates the
spin direction away from the Fermi level.}
\label{FigQAH}
\end{figure}

We first discuss the system at $E_{z}=0$ and compare our results with those
previously obtained in graphene\cite{Qiao,Tse,Yang}. The main difference is
the appearence of almost flat edge modes in our system (Fig.\ref{FigQAH}).
This occurs because the Rashba interactions are different between these two
systems. We have $\lambda _{\text{R1}}=0$ for $E_{z}=0$ and $\lambda _{\text{%
R2}}=0$ at the K and K' points in silicene, but $\lambda _{\text{R1}}\neq 0$
and $\lambda _{\text{R2}}=0$ in graphene. Nevertheless, the difference is
only quantitative. As far as the topological properties are concerned, there
exists no difference. Indeed, in these two systems, the Chern number is
identical in each corresponding phase together with quantized Hall
conductivity, and the edge states support the edge current. However, the
group velocity of the edge modes is extremely small due to the almost flat
gapless modes in silicene.

Our most important result is the VPM phase, which appears in such regions
that $E_{z}M\neq 0$ and occupies a major part of the phase diagram. A part
of the conduction (valence) band is above (below) the Fermi level at the K
(K') point for $E_{z}>0$, as is observed in Fig.\ref{FigRBand}(VPM). Hence,
electrons are moved from the K valley to the K' valley, as implies the
valley polarization. The phase is characterized by the property that it is a
metallic state though gaps are open both at the K and K' points. We note
that the Chern and spin-Chern numbers are ill-defined in the VPM phase,
since the Fermi level does not lie inside the band gaps at the K and K'
points simultaneously.

There exist M-VPM states on phase boundaries indicated by heavy lines in the
phase diagram, where the conduction and valence bands touch the Fermi
surface at the K and K' points, respectively, for $E_{z}>0$. On the other
hand, in SVPM states the conduction and valence bands touch the Fermi
surface both at the K and K' points. We expect topological quantum critical
phenomena in these states. 

In order to explore the physics underlying the phase diagram, we analyze the
low-energy effective Hamiltonian derived from the tight binding model (\ref%
{BasicHamil}). It is described by the Dirac theory around the $K_{\eta }$
point as%
\begin{align}
H_{\eta }=& \hbar v_{\text{F}}\left( \eta k_{x}\tau _{x}+k_{y}\tau
_{y}\right) +\eta \tau _{z}h_{11}+\ell E_{z}\tau _{z}+M\sigma _{z}  \notag \\
& +\lambda _{\text{R1}}(\eta \tau _{x}\sigma _{y}-\tau _{y}\sigma _{x})/2
\label{DiracHamil}
\end{align}%
with $h_{11}=\lambda _{\text{SO}}\sigma _{z}+a\lambda _{\text{R2}}\left(
k_{y}\sigma _{x}-k_{x}\sigma _{y}\right) $, where $\tau _{a}$ is the Pauli
matrix of the sublattice pseudospin, $v_{\text{F}}=\frac{\sqrt{3}}{2}at$ is
the Fermi velocity, and $a=3.86$\AA\ is the lattice constant.

\begin{figure}[t]
\centerline{\includegraphics[width=0.48\textwidth]{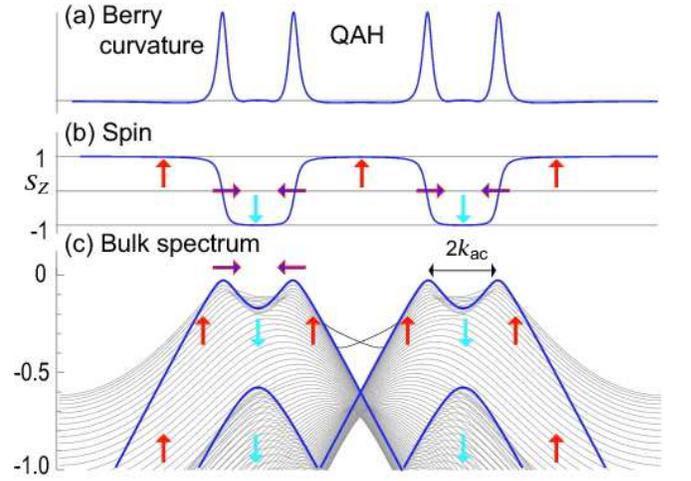}}
\caption{(Color online) (a) Berry curvature, (b) spin, and (c) band
structure of a QAH state calculated based on formula (\protect\ref{SpectM})
in the Dirac theory. Gray curves represent the energy spectrum of a
nanoribbon, which are identical to Fig.\protect\ref{FigQAH}. Spins rotate by
the Rashba interaction near the Fermi level, generating a skyrmion spin
texture in the momentum space. It generates a nontrivial Berry curvature
along the anticrossing circle whose radius given by (\protect\ref{AntiRadiu}%
). The integration of the Berry curvature gives the Chern number $C=2$,
since there are two skyrmions each of which yields $C=1$. }
\label{FigBerryQAH}
\end{figure}

The Hamiltonian $H_{+}$\ explicitly reads%
\begin{equation}
\left( 
\begin{array}{cccc}
E(1,1) & \hbar v_{\text{F}}k_{-} & ia\lambda _{\text{R2}}k_{-} & 0 \\ 
\hbar v_{\text{F}}k_{+} & E(1,-1) & -i\lambda _{\text{R1}} & -ia\lambda _{%
\text{R2}}k_{-} \\ 
-ia\lambda _{\text{R2}}k_{+} & i\lambda _{\text{R1}} & E(-1,1) & \hbar v_{%
\text{F}}k_{-} \\ 
0 & ia\lambda _{\text{R2}}k_{+} & \hbar v_{\text{F}}k_{+} & E(-1,-1)%
\end{array}%
\right)  \label{HamilExpli}
\end{equation}%
in the basis $\left\{ \psi _{A\uparrow },\psi _{B\uparrow },\psi
_{A\downarrow },\psi _{B\downarrow }\right\} ^{t}$, where $k_{\pm }=k_{x}\pm
ik_{y}$, and the diagonal elements are%
\begin{equation}
E(s_{z},t_{z})=\lambda _{\text{SO}}s_{z}t_{z}+\ell E_{z}t_{z}+Ms_{z},
\label{DiagoE}
\end{equation}%
with the spin $s_{z}=\pm 1$ and the sublattice pseudospin $t_{z}=\pm 1$.
They are not good quantum numbers in general. However, since $\lambda _{%
\text{R1}}$ and $\lambda _{\text{R2}}$ are very small with respect to the
other parameters, it is a good approximation to set $\lambda _{\text{R1}%
}=\lambda _{\text{R2}}=0$ in most cases. Thus the spin $s_{z}$ is almost a
good quantum number in general. An exceptional case occurs when two Dirac
cones collapse and cross each other, forming a QAH state after taking into
account the effect of $\lambda _{\text{R2}}\neq 0$, as we soon discuss.

We diagonalize the Hamiltonian (\ref{HamilExpli}) and obtain four energy
levels. When two energy levels coincide, the band gap becomes zero, as found
in Fig.\ref{FigRBand}(M,VMP3,VMP2,SVPM). This occurs at the K and K' points,
where $k_{\pm }=0$. Let us temporarily neglect $\lambda _{\text{R1}}$
because it is very small. Then, the band closes when $%
E(s_{z},t_{z})=E(s_{z}^{\prime },t_{z}^{\prime })$ with (\ref{DiagoE}). They
yield four lines described by $E_{z}=\pm \lambda _{\text{SO}}/\ell $ for $%
\left\vert M\right\vert \leq \lambda _{\text{SO}}$, $M=\pm \lambda _{\text{SO%
}}$ for $\left\vert E_{z}\right\vert \leq \lambda _{\text{SO}}/\ell $, and
two lines by $M=\pm (\ell \lambda _{\text{SO}}/\lambda _{\text{SO}})E_{z}$
outside the square. They are illustrated by dotted lines in Fig.\ref%
{FigPhaseDB}. These lines are modified by the nonzero effect of $\lambda _{%
\text{R1}}$, but the modification is too small to be recognized in Fig.\ref%
{FigPhaseDB}. See also (\ref{StepA}) for the typical order of correction.

The Hamiltonian can be diagonalized analytically in some cases. First, along
the $E_{z}$-axis in the\ phase diagram [Fig.\ref{FigPhaseDB}], we have
already demonstrated\cite{EzawaNJP} that a topological phase transition
occurs along the $E_{z}$-axis from the QSH insulator [Fig.\ref{FigRBand}%
(QSH)] to the band insulator [Fig.\ref{FigRBand}(BI)]. The critical point is
given by%
\begin{equation}
E_{\text{c}}=\pm \frac{2\lambda _{\text{SO}}}{\ell }\left[ \frac{\sqrt{%
1+(\alpha /\ell )^{2}}-1}{(\alpha /\ell )^{2}}\right] ,  \label{StepA}
\end{equation}%
where we have set $\lambda _{\text{R1}}(E_{z})=\alpha E_{z}$ with $\alpha
=10^{-3}$\r{A}. Note that the effect of $\lambda _{\text{R1}}$ is negligible,
$(\alpha /\ell )^{2}=10^{-4}$. The SVPM realizes
at the critical point, where helical currents flow in the bulk.

Second, along the $M$-axis, the first Rashba interaction vanishes ($\lambda
_{\text{R1}}=0$), and the energy spectrum reads%
\begin{equation}
\mathcal{E}=\pm \sqrt{a^{2}\lambda _{\text{R2}}^{2}k^{2}+\left( M-s\sqrt{%
\lambda _{\text{SO}}^{2}+\hbar ^{2}v_{\text{F}}^{2}k^{2}}\right) ^{2}}.
\label{SpectM}
\end{equation}%
We study a topological phase transition along the $M$-axis based on this
formula (\ref{SpectM}). When $M=0$, there are two spin-degenerate Dirac
cones for conduction and valence bands with a gap between them [Fig.\ref%
{FigRBand}(QSH1)]. As $M$ increases, the spin-up (spin-down) Dirac cones are
pushed upward (downward) [Fig.\ref{FigRBand}(QSH2)]. When $\left\vert
M\right\vert \leq \lambda _{\text{SO}}\left( 1+a^{2}\lambda _{\text{R2}%
}^{2}/\hbar ^{2}v_{\text{F}}^{2}\right) $, the band gap is given as $\Delta
=|M-s\lambda _{\text{SO}}|$ at $k=0$, and it closes at $M=s\lambda _{\text{SO%
}}$: This is a topological phase transition point [Fig.\ref{FigRBand}(M)].
Let us temporally assume $\lambda _{\text{R2}}=0$. Then, as $|M|$ increases
further, the two Dirac cones cross each other making a circle around each K$%
_{\mu }$ point. Actually, the Rashba interaction ($\lambda _{\text{R2}}\neq
0 $) mixes up and down spins, turning the crossing points into the
anticrossing points, and opens a gap to form the QAH insulating state [Fig.%
\ref{FigRBand}(QAH) and Fig.3].

When $\left\vert M\right\vert >\lambda _{\text{SO}}\left( 1+a^{2}\lambda _{%
\text{R2}}^{2}/\hbar ^{2}v_{\text{F}}^{2}\right) $, the gap is given by%
\begin{equation}
\Delta =a\lambda _{\text{R2}}\sqrt{\frac{M^{2}}{\hbar ^{2}v_{\text{F}%
}^{2}+a^{2}\lambda _{\text{R2}}^{2}}-\frac{\lambda _{\text{SO}}^{2}}{\hbar
^{2}v_{\text{F}}^{2}}}  \label{GapQAH}
\end{equation}%
at%
\begin{equation}
k_{\text{ac}}=\frac{\sqrt{\hbar ^{4}v_{\text{F}}^{4}\left( M^{2}-\lambda _{%
\text{SO}}^{2}\right) -a^{2}\lambda _{\text{R2}}^{2}\lambda _{\text{SO}%
}^{2}(2\hbar ^{2}v_{\text{F}}^{2}+a^{2}\lambda _{\text{R2}}^{2})}}{\hbar v_{%
\text{F}}\left( \hbar ^{2}v_{\text{F}}^{2}+a^{2}\lambda _{\text{R2}%
}^{2}\right) }.  \label{AntiRadiu}
\end{equation}%
We present the energy spectrum (\ref{SpectM}) and the Berry curvature
calculated by using the corresponding wave function at $M=2\lambda _{\text{SO%
}}$ in Fig.\ref{FigBerryQAH}. As explained there, spins rotates across the
anticrossing point, generating a skyrmion spin texture in the momentum
space. This is consistent with the previous study for graphene\cite{Tse}.
The radius of the anticrossing circles is given by (\ref{AntiRadiu}). We
comment that the gap (\ref{GapQAH}) is of the order of $\mu $eV when $M$ is
of the order of meV.

We now examine a point in the phase diagram such that $ME_{z}\neq 0$. In all
regions where the effects of $\lambda _{\text{R1}}$ and $\lambda _{\text{R2}%
} $ are negligible, the energy spectrum is derived as%
\begin{equation}
\mathcal{E}=s_{z}M\pm \sqrt{\hbar ^{2}v_{\text{F}}^{2}k^{2}+\left( \ell
E_{z}-\eta s_{z}\lambda _{\text{SO}}\right) ^{2}}.  \label{SpectR}
\end{equation}%
The effect of $E_{z}$ is to change the mass of the Dirac electron. Let us
increase $E_{z}$ from $E_{z}=0$ at a fixed value of $M$. The mass decreases
(increases) for the Dirac cone characterized by $\eta s_{z}=+1$ ($\eta
s_{z}=-1$) until $E_{z}=\lambda _{\text{SO}}/\ell $, but the behavior
becomes opposite after $E_{z}=\lambda _{\text{SO}}/\ell $. As a result the
tip of each Dirac cone is pushed either downward or upward as indicated in
Fig.\ref{FigRBand}. Consequently the valley symmetry is broken. Note that
the energy difference at each momentum $\hbar k$ between the conduction and
valence bands with the same spin is given by%
\begin{equation}
\Delta \mathcal{E}_{\pm }=2\sqrt{\hbar ^{2}v_{\text{F}}^{2}k^{2}+\left( \ell
E_{z}\mp \lambda _{\text{SO}}\right) ^{2}}
\end{equation}%
for $\eta s_{z}=\pm 1$, and this is independent of $M$. Thus, the difference
is smaller for the up-spin Dirac cones at the K point, but this is opposite
at the K' point.

We finally determine the phase boundary. It is determined as a boundary
between insulating and metallic states. The Chern and spin-Chern numbers are
quantized in insulating states, while they are ill-defined in metallic
states. As we have seen, each Dirac cone moves upward or downward oppositely
at the K and K' points. Because of this phenomenon the system can become
metallic though the gap is open both at the K and K' points. This is the VPM
state. It occurs when one valence band crosses the Fermi level. The
condition yields four heavy lines ($M=\pm \lambda _{\text{so}}\pm \ell E_{z}$%
) in the phase diagram (Fig.\ref{FigPhaseDB}). On the other hand, the gap
formula (\ref{GapQAH}) determines the boundary between the QAH phase and the
VPM phase, which are the parabolic curves in the phase diagram (Fig.\ref%
{FigPhaseDB}).
In passing we comment that the VPM phase is metallic in nature and does not have mobility gap.
Thus the transition from insulator to VPM might accompany a mobility gap closing.

I am very much grateful to N. Nagaosa for many fruitful discussions on the
subject. This work was supported in part by Grants-in-Aid for Scientific
Research from the Ministry of Education, Science, Sports and Culture No.
22740196.

\end{document}